\documentclass[epj]{svjour}

\usepackage{graphics,epsfig,euscript}
\def\nslash{\rlap{\hspace{0.02cm}/}{n}}
\def\vslash{\rlap{\hspace{0.02cm}/}{v}}
\def\F{{\EuScript F}}

\begin{document}

\title{\boldmath The renormalized B-meson light-cone 
distribution amplitude}

\author{Bjorn O.~Lange}

\institute{
Institute for High-Energy Phenomenology,
Newman Laboratory for Elementary-Particle Physics, \\
Cornell University, Ithaca, NY 14853, U.S.A. \hfill {\tt CLNS 03/1850}}

\date{Received: date / Revised version: date}

\abstract{
I discuss the renormalization-group equation governing the leading order 
light-cone distribution amplitude of the B-meson $\phi^B_+(\omega,\mu)$ and 
its exact analytic solution. The solution displays two features concerning the 
asymptotic behaviour of $\phi^B_+(\omega,\mu)$ for small and large values of 
$\omega$. I comment on further applications and argue that the loss of 
normalizability is not a problem in practice.
} 

\authorrunning{Bjorn O.~Lange}
\titlerunning{The renormalized B-meson light-cone distribution amplitude}

\maketitle

\section{Introduction}
The computation of many decay amplitudes of the $B$-meson simplifies 
considerably in the framework of factorization \cite{Beneke:1999br}, in which 
amplitudes can be expressed in leading power as a convolution integral of 
a calculable hard-scattering kernel and leading order light-cone distribution 
amplitudes (LCDAs) of the mesons involved. Two such amplitudes appear in the 
parameterization of $B$-meson matrix elements of non-local operators where the 
non-locality is light-like. In most applications only one of them, called 
$\phi^B_+(\omega,\mu)$, contributes at leading power and can be defined as the 
Fourier transform of $\widetilde\phi_+^B(\tau,\mu)$, where \cite{Grozin:1996pq}
\begin{eqnarray}\label{LCDA}
   &&\langle\,0\,|\,\bar q_s(z)\,S_n(z,0)\,\nslash\,\Gamma\,h(0)\,
    |\bar B(v)\rangle \nonumber\\[0.3cm]
   &&\qquad = - \frac{iF(\mu)}{2}\,\widetilde\phi_+^B(\tau,\mu)\,
    \mbox{tr}\left( \nslash\,\Gamma\,\frac{1+\vslash}{2}\,\gamma_5
    \right) .
\end{eqnarray}

A sensitivity to this universal, non-perturbative function arises in processes 
which involve hard interactions with the soft spectator quark in the $B$-meson,
for example in $B \to \pi l \nu$, $B \to K^* \gamma$, etc. A generic 
factorizable decay amplitude may be written as 
\begin{equation}\label{conv}
  {\mathcal A}=\int_0^\infty\frac{d\omega}{\omega}\,T(\omega,\mu)\,
   \phi_+^B(\omega,\mu) \,,
\end{equation}
where we assume that the $\mu$ dependence cancels between the hard-scattering 
kernel $T$ and the LCDA. In writing the amplitude in this way we have achieved 
a first step of scale separation, since the kernel depends on the physics 
associated with large energy scales, i.e. does not contain large logarithms for
$\mu \sim m_b$, whereas the LCDA is a universal, non-perturbative function 
which ``lives'' on low scales $\mu \sim \Lambda_{QCD}$. Since there is no one 
scale at which neither of the two quantities contain large logarithms it is 
crucial to resum those large (Sudakov) logarithms to gain full control over 
the separation of physics at different scales. We thus have to derive and 
solve the renormalization group evolution for the LCDA or, equivalently, the 
hard-scattering kernel \cite{Lange:2003ff,Bosch:2003fc}. In this 
(admittedly technical) talk I present the renormalization group evolution 
equation of the LCDA $\phi_+^B(\omega, \mu)$, its exact solution, and derive 
scaling properties for its asymptotic behaviour.

\section{Derivation of the RGE}

Since different operators that share the same quantum numbers can mix under 
renormalization we write the relation between bare and renormalized 
operators as
\begin{equation}
O_+^{ren}(\omega,\mu) = 
\int d\omega'\; Z_+(\omega, \omega',\mu) \; O_+^{bare}(\omega') \; .
\label{eq:op}
\end{equation}

In the case at hand the operator is, up to a Dirac trace, the product of 
$F(\mu)$, which denotes asymptotic value of $\sqrt{m_B} f_B$ in the 
heavy-quark limit, and the LCDA $\phi_+^B(\omega,\mu)$. The analytic 
structure of $\widetilde \phi^B_+(\tau,\mu)$ implies that 
$\phi_+^B(\omega,\mu)$ vanishes for negative $\omega$. It then follows that 
the Renormalization Group Equation is an integro-differential equation in which
the anomalous dimension is convoluted with the LCDA.
\begin{equation}
\frac{d}{d\ln \mu} \phi_+^B(\omega,\mu) = 
- \int\limits_0^\infty d\omega' \; \gamma_+(\omega, \omega',\mu) \;
  \phi_+^B(\omega',\mu)
\label{eq:RGE}
\end{equation}

\begin{figure}[t!]
\begin{center}
\epsfig{file=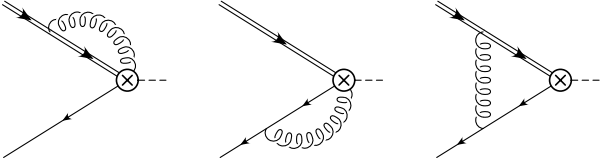, width=8cm}
\caption{One-loop diagrams for the calculation of the anomalous dimension of 
the $B$-meson light-cone distribution amplitude.}
\label{fig:oneloop}
\end{center}
\end{figure}

We may separate the on- and off-diagonal terms of the anomalous dimension and 
express it as
\begin{eqnarray}
\gamma_+(\omega, \omega',\mu) &=& 
\Big[ \Gamma_{cusp}(\alpha_s) \ln \frac{\mu}{\omega} 
     + \gamma(\alpha_s) \Big] \delta(\omega-\omega') \nonumber \\ 
&& + \omega \; \Gamma(\omega, \omega', \alpha_s) 
\label{eq:anomdim}
\end{eqnarray}
to all orders in perturbation theory. The cusp anomalous dimension 
$\Gamma_{cusp}(\alpha_s)$ appears as the coefficient of the $\ln \mu$ term 
and has a geometric origin \cite{Korchemsky:wg}: 
Since an effective heavy-quark field $h(0)$ can be expressed as the product 
of a free field and a Wilson line $S_v(0,-\infty)$ extending from $(-\infty)$ 
to $0$ along the $v$-direction, the matrix element in (\ref{LCDA}) contains 
$S_n(z,0) \,S_v(0,-\infty)$ which can be combined to form a single Wilson 
line with a cusp at the origin, as shown in Fig. \ref{fig:2}(a). 
The appearance of the single $\ln \mu$ term distinguishes the anomalous 
dimension of the $B$-meson LCDA from the familiar Brodsky-Lepage kernel 
\cite{Lepage:1980fj}. On the one-loop level the $\ln \mu$ term appears in 
the calculation of the first diagram in Fig. \ref{fig:oneloop}, 
where the gluon from the Wilson line $S_n(z,0)$ connects to the heavy-quark 
Wilson line $S_v(0,-\infty)$. 

We find the one-loop expressions \cite{Lange:2003ff} (denoted by the 
superscript $^{(1)}$) to be $\Gamma_{cusp}^{(1)}=4$, $\gamma^{(1)}=-2$, and
\begin{equation}
\Gamma^{(1)}(\omega, \omega') = - \Gamma_{cusp}^{(1)} \left[
  \frac{\theta(\omega'-\omega)}{\omega' (\omega'-\omega)}
+ \frac{\theta(\omega-\omega')}{\omega (\omega-\omega')} \right]_+ \; ,
\label{eq:1-loop}
\end{equation}
where we have used that $\gamma_F^{(1)}=-3$ is the one-loop coefficient of 
the anomalous dimension of heavy-to-light currents. The subscript $+$ denotes 
the standard ``plus distribution'' which ensures that 
$\int d\omega' \Gamma(\omega, \omega') = 0$.

\section{Keys to the exact solution}

The first key toward solving the RG Eq. (\ref{eq:RGE}) concerns the 
off-diagonal term $\omega \Gamma(\omega, \omega',\mu)$ in the anomalous 
dimension (\ref{eq:anomdim}). We observe that (omitting $\alpha_s \equiv 
\alpha_s(\mu)$ dependence)
\begin{equation}
\int\limits_0^\infty d\omega' \omega \Gamma(\omega, \omega') (\omega')^a
= \omega^a \F(a)
\label{eq:F}
\end{equation}
on dimensional grounds. The dimensionless function $\F$ can only depend on the 
(in general complex) exponent $a(\mu)$, which in turn must be allowed to 
depend on the renormalization scale $\mu$. We can therefore use a power-law 
ansatz 
\begin{equation}
f(\omega,\mu,\mu_0,a(\mu)) = \left( \frac{\omega}{\mu_0} \right)^{a(\mu)} \; 
                              e^{U(a(\mu),\mu)}
\end{equation}
with an arbitrary mass parameter $\mu_0$. The function $f$ solves the RG Eq. 
(\ref{eq:RGE}), if the exponent $a(\mu)$ and the normalization $U(a(\mu),\mu)$
obey the differential equations 
\begin{eqnarray}
\frac{d}{d\ln \mu} \, a(\mu) &=& \Gamma_{cusp}(\alpha_s) \; , \\
\frac{d\, U(a(\mu),\mu)}{d\ln \mu} \, &=& -\gamma(\alpha_s) 
  - \F(a(\mu),\alpha_s) - \ln \frac{\mu}{\mu_0} \, 
  \Gamma_{cusp}(\alpha_s) \; . \nonumber
\end{eqnarray}
The first equation can be immediately integrated and yields $a(\mu) = \eta + 
g(\mu,\mu_0)$ with initial value $\eta = a(\mu_0)$ and 
\begin{equation}
g(\mu,\mu_0) = \int\limits_{\alpha_s(\mu_0)}^{\alpha_s(\mu)} 
         \frac{d\alpha}{\beta(\alpha)} \; \Gamma_{cusp}(\alpha) \; .
\label{eq:g}
\end{equation}
With this solution at hand, the second equation integrates to
\begin{eqnarray}
U(a(\mu),\mu)&=& - \int\limits_{\alpha_s(\mu_0)}^{\alpha_s(\mu)} 
         \frac{d\alpha}{\beta(\alpha)} \; \Big[ 
         \gamma(\alpha) + g_\mu(\alpha) \nonumber \\
 && \hspace{20mm} + \F(\eta+ g_0(\alpha), \alpha) \Big] \; ,
\label{eq:U}
\end{eqnarray}
where $g_\mu(\alpha)=g(\mu,\mu_\alpha)$, $g_0(\alpha)=g(\mu_\alpha, \mu_0)$, 
and $\mu_\alpha$ is defined such that $\alpha_s(\mu_\alpha) = \alpha$. Note 
that $g(\mu_0,\mu_0)=0$ and $U(\eta,\mu_0)=0$ in this construction.
\\

The second key to the solution concerns the initial condition 
$\phi_+^B(\omega, \mu_0)$. Defining the Fourier transform $\varphi_0(t)$ of 
the LCDA at scale $\mu_0$ with respect to $\ln (\omega/\mu_0)$ allows us to 
express the $\omega$ dependence in the desired power-law form
\begin{equation}
\phi_+^B(\omega, \mu_0) = \frac{1}{2\pi} \int\limits_{-\infty}^\infty dt \; 
             \varphi_0(t) \left( \frac{\omega}{\mu_0} \right)^{it} \; .
\end{equation}
We therefore obtain an exact analytic expression for the solution of the RG 
Eq. (\ref{eq:RGE}) as the single integral
\begin{equation}
\phi_+^B(\omega, \mu) = \frac{1}{2\pi} \int\limits_{-\infty}^\infty dt \; 
             \varphi_0(t) \; f(\omega, \mu, \mu_0, \eta=it) \; .
\label{eq:sol}
\end{equation}

\section{Asymptotic behaviour}

\begin{figure}[b!]
\begin{center}
{\bf (a)} \epsfig{file=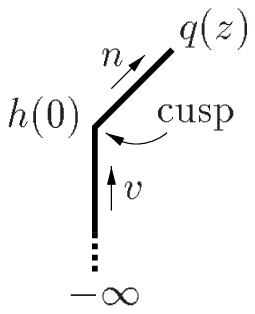, height=25mm} \hspace{10mm} 
{\bf (b)} \epsfig{file=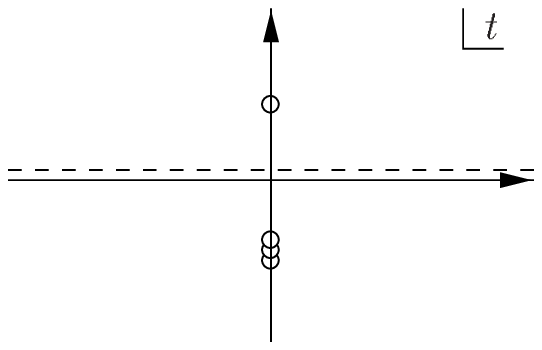, height=25mm}
\caption{{\bf (a)} Left: The cusp in the Wilson line $S_n(z,0) S_v(0,-\infty)$.
{\bf (b)} Right: Poles of the function $f(\ldots,it)$ in the complex $t$ plane.
The upper pole remains stationary under renormalization flow, whereas the 
position of the lower pole moves toward the real axis for increasing $\mu$.}
\label{fig:2}
\end{center}
\end{figure}

The solution (\ref{eq:sol}) enables us to extract the asymptotic behaviour of 
the LCDA $\phi_+^B(\omega,\mu)$ as $\omega \to 0$  and $\omega \to \infty$ by 
deforming the integration contour in the complex $t$ plane. We hence need 
to study the analytic structure of the integrand $\varphi_0(t) \; 
f(\omega, \mu, \mu_0, it) \sim \omega^{it + g(\mu,\mu_0)}$. If $\omega$ is 
very small we can deform the contour into the lower half plane and then the 
position of the nearest pole to the real axis determines the $\omega$ 
dependence of $\phi_+^B(\omega,\mu)$. Similarly the nearest pole in the upper 
half plane dominates for very large $\omega$.

Let us study the analytic structure of $f(\omega, \mu, \mu_0, it)$ at leading 
order in RG-improved perturbation theory. Using the one-loop expressions 
(\ref{eq:1-loop}) and the definition (\ref{eq:F}) we find 
\begin{equation}
\F^{(1)}(a) = \Gamma_{cusp}^{(1)} \left[ \Psi(1+a) + \Psi(1-a) + 
                          2 \gamma_E \right] \; .
\end{equation}
$\Psi$ and $\gamma_E$ denote the logarithmic derivative of the Euler-Gamma 
function and the Euler-constant, respectively. Plugging this result into 
eq. (\ref{eq:U}) with $\eta = it$ one obtains (using the short-hand notation 
$g \equiv g(\mu,\mu_0)$)
\begin{equation}
e^{U(it+g,\mu)} \propto 
\frac{\Gamma(1+it)\; \Gamma(1-it-g)}{\Gamma(1-it)\; \Gamma(1+it+g)} \; .
\end{equation}
The function $f(\omega, \mu, \mu_0, it)$ has poles along the imaginary axis, 
and the closest to the real axis are located at $t=i$ and $t=-i(1-g)$. Using 
the one-loop expression (\ref{eq:g}) we observe that the function $g$ vanishes 
at $\mu=\mu_0$ by definition and grows monotonously as $\mu$ increases. 
Therefore the position of the pole in the lower complex plane approaches the 
real axis under renormalization evolution, as illustrated in 
Fig. \ref{fig:2}(b). 

These poles ``compete'' with the singularities arising from $\varphi_0(t)$ 
for the nearest position to the real axis. Let us assume that, for a given 
model of $\phi_+^B(\omega, \mu_0)$, the LCDA grows like $\omega^\delta$ for 
small $\omega$ and falls off like $\omega^{-\xi}$ for large $\omega$. The 
corresponding poles of the function $\varphi_0(t)$ are then located at 
$t=-i\delta$ and $t=i\xi$. We therefore obtain the asymptotic behaviour of 
the renormalized LCDA as
\begin{equation}
   \phi_+^B(\omega,\mu)\sim
   \cases{ \omega^{\min(1,\delta+g)} \,; & for $\omega\to 0$, \cr
           \omega^{-\min(1,\xi)+g} \,; & for $\omega\to\infty$. \cr}
\label{eq:asympt}
\end{equation}
The two immediate observations are that, regardless of how small the value 
of $\delta$ is, evolution effects will drive the small $\omega$ behaviour 
toward linear\footnote{Might this be an argument for $\delta=1$ in the first 
place?} growth, and that the renormalized LCDA at a scale $\mu>\mu_0$ 
will fall off slower than $1/\omega$ irrespective of how fast it vanishes at 
$\mu=\mu_0$.

\section{Concluding Remarks}

The emergence of a radiative tail after (even infinitesimally small) 
evolution seems, at first sight, a very strange property of the LCDA, because 
it implies that the normalization integral of $\phi_+^B(\omega,\mu)$ is UV 
divergent. This can be understood as the corresponding {\em local} operator 
$\tilde \phi_+^B(\tau=0,\mu)$ requires an additional subtraction when 
renormalized. However, this is not an obstacle in practical applications in 
which only $\phi_+^B(\omega,\mu)/\omega$ modulo logarithms appears. 
An integral over this function remains UV finite as long as $g(\mu,\mu_0)<1$, 
at which point the pole at $t=-i(1-g)$ reaches the real axis and the above 
formalism breaks down. 

It is evident from (\ref{eq:asympt}) that evolution effects mix different 
moments of the LCDA. For example, the first inverse moment of 
$\phi_+^B(\omega, \mu)$ defines a parameter $\lambda_B(\mu)$,
\begin{equation}
\frac{1}{\lambda_B(\mu)} = \int\limits_0^\infty \frac{d\omega}{\omega} \; 
                 \phi_+^B(\omega,\mu) \; ,
\end{equation}
which is connected to a fractional inverse moment of order $1-g(\mu,\mu')$ at 
a different scale $\mu'$. This makes it impossible to calculate the scale 
dependence of $\lambda_B(\mu)$ in perturbation theory without knowledge of 
the LCDA. 
\\

In this talk I presented the renormalization-group equation of the $B$-meson 
LCDA and discussed the key steps in solving this integro-differential equation 
analytically. Since the amplitude (\ref{conv}) is independent of the 
renormalization scale $\mu$, the technique presented here can be used to resum 
Sudakov logarithms in the hard scattering kernel $T(\omega, \mu)$, which was 
demonstrated in reference \cite{Bosch:2003fc} for the $B \to \gamma l \nu$ 
decay mode. 

Let me stress that the solution outlined here can also be used in other 
applications in $B$-physics, such as the the inclusive $B \to X_s \gamma$ for 
example, in which the shape-function and the hard-scattering kernel relevant 
to the analysis of the photon spectrum \cite{Neubert:1993um,Bauer:2000ew} obey 
evolution equations similar to (\ref{eq:RGE}).

\subsection*{Acknowledgments}
I would like to thank T. Mannel for lunch and an interesting discussion.

\end{document}